\documentstyle[multicol,prb,aps]{revtex}

\draft

\begin{document}
\title{Fractal approach to the $\beta$ relaxation in supercooled liquids}
\author{Alexei V\'{a}zquez$^{1}$, Oscar Sotolongo-Costa$^{1,2}$}
\address{1-Department of Theoretical Physics. Havana University. Havana 10400, Cuba.\\
2- Departamento de Fisica-Matematica y Fluidos. Fac. Ciencias. UNED. A.P\\
60141. 28080 Madrid, Spain.}
\author{Francois Brouers}
\address{Institute of Physics, Li\`ege University, B5, 4000 Li\`ege, Belgium.}
\date{\today}
\maketitle

\begin{abstract}
In the present work we present a fractal model for the $\beta$ relaxation in
supercooled liquids. The macroscopic dynamics is obtained by superposition
of relaxation of independent mesoscopic regions (cages). Scaling relations
and exponents are assumed for the distribution of cage sizes and for the
size dependent response of the independent cages. In this way we obtain some
scaling relations for the average response, in the time and frequency
domain. For a particular choice of the scaling exponents we obtain the
scaling relations predicted by the MCT. The comparison with recent light
scattering data reveals that the scaling exponents of the distribution of
cage sizes are universal but the single cage relaxation depends on the
detailed material structure.
\end{abstract}

\pacs{61.20.Lc, 64.70.Pf, 61.43.H}

\section{Introduction}

The liquid-glass transition at the calorimetric temperature $T_g$ has been a
subject of extensive experimental and theoretical study for many years.
However, more recently the interest has been shifted to the range of
supercooled liquid temperatures, which is extended from $T_g$ to the melting
temperature $T_m$. Light-scattering experiments in this range of
temperatures reveals two relaxation processes, $\alpha$ and $\beta$
relaxations \cite{review,li2,cummins,sidebottom2,wuttke,li1,sidebottom1} .

The commonly accepted explanation for the $\beta$ relaxation is that it is
related to the molecular details of the glass former. However, the mode
coupling theory (MCT) \cite{MCT} suggests that it may be a general dynamical
feature of dense disordered systems rather than being specific to the glass
under investigation. Moreover, the MCT theoretically demonstrates the
existence of the two distinct relaxation processes. The slower process ($%
\alpha$) is associated with the diffusive motion and the faster one ($\beta$%
) is attributed to strongly anharmonic motions of a limited number of
molecules trapped in cages formed by their neighbors.

On the other hand, computer simulations suggest that the $\beta$ relaxation
represents cooperative localized motions of a small number of molecules
within a cage formed by neighboring molecules \cite{wahnstrom}. On the time
scale of the $\beta$ relaxation , these cages appear frozen. For longer
times collective rearrangements continue allowing the eventual relaxation of
the cages, which represents the $\alpha$ relaxation.

If one neglects the interaction between different cages then one may model
the $\beta$ relaxation as the superposition of a heterogeneous distribution
of independent relaxing regions, the cages. There are several models in the
literature which deal with such a situation \cite
{domb,cohen,vazquez,chamberlin}. In all cases, it is assumed that each
cluster relaxes exponentially with a size dependent relaxation time. The
main simplification of these models is thus the assumption of an exponential
relaxation of the independent regions. The slow relaxation dynamics appears
when one superposes the contribution of each region over the distribution of
cluster sizes. However, the cooperative nature of the dynamics inside each
cluster leads, by itself, to a slow relaxation dynamics. In fact, there are
several models which do not contain disorder and obtain the slow relaxation
as a consequence of the cooperative nature of the dynamics \cite
{MCT,ngai,franz}.

In the present work we present a fractal model to describe the main features
of the $\beta$ relaxation in supercooled liquids. We assume that the
macroscopic dynamics is given by the superposition of the relaxation of
independent mesoscopic regions (cages). Nevertheless, different from
previous works, we do not assume that each mesoscopic region relaxes
exponentially. In this way we obtain some scaling relations for the average
response, in the time and frequency domain. For a particular choice of the
scaling exponents we obtain the scaling relations predicted by the MCT. The
comparison with recent light scattering data reveals that the scaling
exponents of the distribution of cage sizes are universal but the single
cage relaxation depends on the detailed material structure.

The paper is organized as follows. In section \ref{sec:model} we introduce
the fractal model for the $\beta$ relaxation in supercooled liquids and
derive some of its predictions. The scaling relations for the normalized
relaxation function and the imaginary part of the susceptibility are
obtained. Then, in section \ref{sec:discussion} we discuss the main results
of the model, and compare them with other theoretical approaches and recent
light-scattering data reported in the literature. Finally, the conclusions
are given in section \ref{sec:conclusions}.

\section{Fractal model}

\label{sec:model}

Computer simulations suggest that the $\beta$ relaxation represents
cooperative localized motions of a small number of molecules within a cage
formed by neighboring molecules \cite{wahnstrom}. On the time scale of the $%
\beta$ relaxation, these cages appear frozen. For longer times collective
rearrangements continue allowing the eventual relaxation of the cages, which
represents the $\alpha$ relaxation. In the following we restrict our
analysis to the mesoscopic time scale between the microscopic characteristic
time $\tau_0$ and the mean live time of the cages, i.e. we focus our
attention in the $\beta$ relaxation regime.

\subsection{Single cage relaxation}

Let $s$ be the number of molecules in a cage and $\tau_s$ the characteristic
time associated with the cooperative rearrangements of the molecules in a
cage of size $s$. The relaxation in large cages, involving many molecules,
is relatively infrequent and is associated with a long time relaxation
process. On the contrary, the relaxation in small cages, involving a few
molecules, occurs more often and is associated with a short time relaxation
process. We thus assume that the characteristic time scales with the cage
size according to 
\begin{equation}
\tau_s=\tau_0 s^\zeta,  \label{eq:1}
\end{equation}
where $\zeta>0$ is a scaling exponent. In the following we measure time in
units of $\tau_0$.

$\tau_s$ is the only relevant time scale in the $\beta$ relaxation regime.
Hence, the normalized response function will satisfy the scaling law 
\begin{equation}
\Phi(s,t) = t^\eta f_1(s^{-{\zeta}}t).  \label{eq:2}
\end{equation}
where $f_1(x)$ is a scaling function. On the other hand, $\Phi(s,t)$ has to
satisfy the normalization condition 
\begin{equation}
\int_0^\infty dt \Phi(s,t) = 1.  \label{eq:2a}
\end{equation}
This constraint carry as a consequence that 
\begin{eqnarray}
\eta = -1 & \hskip 0.2in & \int_0^\infty dx \frac{f_1(x)}{x} =1.
\label{eq:2b}
\end{eqnarray}

\subsection{Average relaxation}

The cage size follows certain distribution $P(s)$. Computer simulations
suggest that the number of molecules involved in cooperative rearrangements
is widely distributed \cite{ohmine,miyagawa}. Let $s_c\sim\epsilon^{-1/%
\sigma}$ be a characteristic cage size below with the distribution is scale
invariant. We can therefore assume that $P(s)$ satisfies the scaling law 
\begin{equation}
P(s) = s^{-\tau}f_2(\epsilon^{1/\sigma}s),  \label{eq:3}
\end{equation}
where $1<\tau<2$ and $\sigma>0$ are scaling exponents, and $f_2(x)$ is a
scaling function with the asymptotic behaviors: $f_2(x\ll 1)\approx \text{%
const.}$ and $f_2(x\gg 1)\ll 1$. $\epsilon$ is a control parameter, which
characterizes how close is the system from the scale invariant case: $%
P(s)\sim s^{-\tau}$. It can be related to the mean cluster size through the
equation 
\begin{equation}
\langle s\rangle \sim \epsilon^{-(2-\tau)/\sigma}.  \label{eq:4}
\end{equation}
Hence $\langle s\rangle\rightarrow\infty$ is equivalent to $%
\epsilon\rightarrow 0$. In supercooled liquids and amorphous solids there is
no long range order and, therefore, we expect that the mean cluster size is
always finite, i.e. $\epsilon>0$. We are thus considering the relaxation of
scale invariant mesoscopic clusters.

The average normalized response function is given by 
\begin{equation}
\Phi(t) = \int_0^\infty ds P(s) \Phi(s,t).  \label{eq:5}
\end{equation}
Using eqs. (\ref{eq:2}), (\ref{eq:2b}) and (\ref{eq:3}) one obtains 
\begin{equation}
\Phi(t) = t^{-1-a}f_3(\epsilon^{1/\Delta} t),  \label{eq:6}
\end{equation}
where 
\begin{equation}
a=(\tau-1)\zeta^{-1}, \ \ \ \ \ \Delta = \sigma/\zeta,  \label{eq:7}
\end{equation}
\begin{equation}
f_3(x) = \zeta^{-1} \int_0^\infty dy y^{a-1}f_1(y)f_2[(x/y)^{1/\zeta}].
\label{eq:8}
\end{equation}
$f_3(x)$ is another scaling function, expressed as a function of the
previously introduced scaling functions $f_1(x)$ and $f_2(x)$. If $f_2(x)$
is analytical in $x=0$ then, expanding this function around $x=0$, from eq. (%
\ref{eq:8}) one obtains 
\begin{equation}
f_3(x)\approx A - Bx^{1/\zeta},  \label{eq:9}
\end{equation}
where 
\begin{equation}  \label{eq:9a}
A = \zeta^{-1} f_2(0) \int_0^\infty dx x^a \frac{f_1(x)}{x} \\
\end{equation}
\begin{equation}
B = \zeta^{-1} |\dot{f}_2(0)| \int_0^\infty dx x^{a-1/\zeta} \frac{f_1(x)}{x}%
.  \label{eq:9b}
\end{equation}
Here we assume that the function $f_1(x)$ is such that the integrals in eqs.
(\ref{eq:2b}), (\ref{eq:9a}) and (\ref{eq:9b}) converge. Substituting this
asymptotic expansion in eq. (\ref{eq:6}) it results that 
\begin{equation}
\Phi(t) \sim \epsilon^{(1+a)/\Delta} [(\epsilon^{1/\Delta}t)^{-1-a} +
C(\epsilon^{1/\Delta} t)^{-1+b}],  \label{eq:10}
\end{equation}
where 
\begin{equation}
b = \zeta^{-1}-a,  \label{eq:11}
\end{equation}
and $C=B/A$. Eq. (\ref{eq:10}) gives the asymptotic behavior of the response
function in the scale invariant region $t\ll\epsilon^{-1/\Delta}$.

\subsection{Susceptibility}

The imaginary part of the normalized susceptibility $\chi {\prime \prime }$
is related to the normalized response function through the equation 
\begin{equation}
\chi ^{\prime \prime }(\omega )=\text{FT}^{\prime \prime }[\Phi (t)](\omega
)\ ,  \label{eq:12}
\end{equation}
where $\text{FT}^{\prime \prime }$ denotes the imaginary part of the Fourier
transform. Substituting $\Phi (t)$ by its scaling law in eq. (\ref{eq:6})
one obtains 
\begin{equation}
\chi ^{\prime \prime }(\omega )=\omega ^{a}f_{4}(\epsilon ^{-1/\Delta
}\omega ),  \label{eq:13}
\end{equation}
where $f_{4}(x)$ is another scaling function, given by 
\begin{equation}
f_{4}(x)=\int_{0}^{\infty }dy\cos (y)y^{-a-1}f_{3}(y/x).  \label{eq:14}
\end{equation}

Moreover, in the frequency range $\epsilon^{1/\Delta}\ll\omega\ll1$ from
eqs. (\ref{eq:10}) and (\ref{eq:12}) it follows that 
\begin{equation}
\chi^{\prime\prime}(\omega) = \chi^{\prime\prime}_{\text{min}} \Biggl[a\Bigl(%
\frac{\omega}{\omega_{\text{min}}}\Bigr)^a + b\Bigl(\frac{\omega}{\omega_{%
\text{min}}}\Bigr)^{-b}\Biggr],  \label{eq:15}
\end{equation}
where 
\begin{equation}
\chi^{\prime\prime}_{\text{min}} \sim \epsilon^{a/\Delta} \ \ \ \ \ \omega_{%
\text{min}} \sim \epsilon^{1/\Delta}.  \label{eq:16}
\end{equation}
From eq. (\ref{eq:15}) one concludes that the susceptibility has a minimum $%
\chi^{\prime\prime}_{\text{min}}$ at the frequency $\omega_{\text{min}}$.

\section{Discussion}

\label{sec:discussion}

In the previous section we have obtained the scaling laws for the normalized
relaxation function and the imaginary part of the susceptibility. Now, we
proceed to discuss these results and to compare them with other theoretical
approaches and experimental data.

The scaling laws in eqs. (\ref{eq:2}) and (\ref{eq:13}) are very similar to
those obtained by the MCT \cite{MCT}. In fact, if we take 
\begin{equation}
\Delta = 2a,  \label{eq:17}
\end{equation}
then we obtain the scaling laws of the MCT. This equation constitutes a
scaling relation for the scaling exponents $a$ and $\Delta$, which has to be
imposed to obtain the results of the MCT. Our formulation thus result more
general, while the MCT prediction is contained as a particular case.

From eqs. (\ref{eq:7} and (\ref{eq:11}) one can obtain the exponents $\tau$
and $\zeta$, introduced in our fractal model, as a function of $a$ and $b$,
which are usually measured in light scattering experiments in the $\beta$
regime. It results that 
\begin{equation}
\zeta=\frac{1}{a+b}, \ \ \ \ \ \tau=\frac{2a+b}{a+b}.  \label{eq:18}
\end{equation}

Moreover, if one assumes that (\ref{eq:18}) is valid then form eq. (\ref
{eq:7}) we obtain 
\begin{equation}
\sigma=\frac{2a}{a+b}.  \label{eq:19}
\end{equation}
We emphasize again that eq. (\ref{eq:18}), and therefore eq. (\ref{eq:19}),
is valid if one assumes the MCT prediction. Then using some light scattering
reports for different supercooled liquids we have calculated these scaling
exponents. The results are shown in \ref{tab:1} and \ref{tab:2}.

First, we notice that the exponent $\tau$ is larger than the one obtained in
standard percolation (SP) in three dimensions, $\tau_{\text{SP}}=1.15$. One
of the shortcomings of standard percolation theory is that it neglects
frustration, which introduces cancellations and hence the possibility of
weak correlation, even if the interactions are much smaller than the thermal
energy. Thus, standard percolation models are too simple to explain the
behavior of glass forming liquids. This conclusion has been also pointed out
for spin glasses \cite{binder}. The inclusion of frustrated effects in
percolation models is relatively recent \cite{coniglio}. The frustrated
percolation (FP) model suggests that in glass-forming liquids there is a
percolation temperature $T_{\text{FP}}$ above $T_g$, where frustration
effects starts to be manifested. This transition leads to highly correlated
localized clusters.

Thus, we can describe the cage formation in supercooled liquids using a
frustrated percolation model. Unfortunately, simulations in three dimensions
are not yet available. In any case, it is expected that the exponents $\tau$
and $\sigma$ are larger in FP than in SP. Frustration effects inhibit the
formation of long range structures and, therefore, the distribution of
cluster sizes in FP should decay faster than in SP, i.e. $\tau_{\text{SP}%
}<\tau_{\text{FP}}$. Moreover, the temperature $T_c$ may be identified with $%
T_{\text{FP}}$ and $\epsilon=|T-T_{\text{FP}}|$.

The data displayed in tables \ref{tab:1} and \ref{tab:2} has been divided in
different groups attending to the structure of the glass formers. In the
first class we have the two simple molecular glass formers ortophenyl and
salol. These materials have a very simple structure with symmetric ( Van der
Waals like ) interactions among their molecules. They are fragile glass
formers and the scaling exponents $a$ and $b$ in both materials are quite
similar. Molecular glass formers thus show similar static and dynamic
behavior. Frustration may arise when local arrangements of molecules
kinetically prevent all the molecules from reaching ordered close-packed
configurations. Frustration effects have in this case a kinetic origin. In
the second group we find the glass formers glycerol and B$_2$O$_3$. Glycerol
is a fragile glass former which has a hydrogen-bonded network structure. B$%
_2 $O$_3$ is a strong glass former and has a covalent bonded network
structure. Hence, these materials are different in relation with the
temperature dependence of the static viscosity. However, the scaling
exponents $a$ and $b $, which characterize the dynamical properties, are in
both materials quite similar. The difference in the temperature dependence
of the static viscosity can be associated to the strength of the bonds. The
hydrogen bonds are weaker than the covalent ones and, therefore, the network
structure in grycerol is more sensitive to temperature shift than in B$_2$O$%
_3$. The common factor in these glass formers is the existence of network
structures. We thus expect that frustration is mainly determined by the
competition of different interactions generated by the multiple random bond
network configurations. The existence of constrained bond angles or lengths
leads to some optimized network structure, where some bonds will not satisfy
these constrains and become partially broken (frustrated). Frustration
effects have, in this case, a structural origin. Finally, in the third and
fourth group we found the ionic glass former CKN and polymers, respectively.
The exponents $a$ and $b$ are in both groups very different and greater than
those reported for the first two groups. The origin of the frustration
effects in these materials is less clear since we do not understand their
structure well enough.

However, in spite that $a$ and $b$ change from one group to another, the
scaling exponents of the distribution of cages sizes $\tau$ and $\sigma$ are
very similar. We found $\tau\approx 1.34$ and $\sigma\approx 1.69$ (an
exception are the polymers PPG and PET which are discussed below). Thus we
have found a universal behavior for the distribution of cages sizes. On the
contrary, the scaling exponent $\zeta$ results material dependent. This
scaling exponent is related to the single cage relaxation and, therefore,
one may conclude that the single cage relaxation depends on the detailed
structure of the glass former. The universal behavior can only appear when
the single cage relaxation is averaged over the distribution of cages sizes.

In the case of the polymers PPG and PET the exponent $\sigma$ deviates from
that observed in the others glass formers. However, the scaling exponent $%
\tau$ is similar to the one observed in the other glass formers. This
deviation may be associated with the fact that the relation between $\tau$
and the light scattering exponents $a$ and $b$ were obtained using general
considerations, while that for $\sigma$ were obtained assuming that the MCT
is correct. Hence, one may conclude that the MCT cannot be applied to
polymers. Off course, to validate this conclusion a more extensive
experimental analysis is needed.

\section{Summary and Conclusions}

\label{sec:conclusions}

We have presented a fractal approach to the $\beta$ relaxation in
supercooled liquids. The main ingredients of the model are the existence of
scaling laws for the single cage normalized relaxation function and for the
distribution of cage sizes. In this way we have obtained the scaling laws
for the average response, which for a particular choice of scaling exponents
lead to those obtained by the MCT.

We have obtained some scaling relations between the scaling exponents
introduced in the model and those obtained from the fit to light-scattering
data. Using these scaling relations and the reports for some
light-scattering experiments we have found the $\beta$ relaxation is indeed
a general dynamical feature of dense disordered systems, which does not
depend on the specific glass former under investigation.

The scaling exponents distribution of cage sizes appears to be universal.
The material dependence of the exponents $a$ and $b$ obtained from light
scattering data is a consequence of the non-universality of the single cage
relaxation, which depends on the detailed structure of the glass former.
Moreover, the scaling exponent $\tau$ and $\sigma$ were found larger than
the one obtained in standard percolation in three dimensions. This
discrepancy was attributed to frustration effects, and the cage formation
was associated to a frustrated percolation problem.

Our approach reveals that the MCT scaling relations may not be valid for
polymers, but a more extensive analysis of experimental data is needed. The
present work may be used as a guide for further experimental investigations
of the $\beta$ relaxation in supercooled liquids and glasses, in order to
check our results.

\section*{Acknowledgements}

This work was partially supported by the {\em Alma Mater} prize, given by
The University of Havana and by the Direccion General de Investigacion
Cientifica y Tecnica (DGICYT, Spain, Ministerio de Educacion y Cultura). The
financial support by the Vicerrectorado de Centros Asociados of the UNED is
gratefully acknowledged

\begin{table}[tbp]
\begin{tabular}{lccccc}
Glass former & $\Delta$ & $\sigma$ &  &  &  \\ \hline
salol & 0.66 & 0.68 &  &  &  \\ 
orthophenyl & 0.66 & 0.67 &  &  &  \\ \hline
B$_2$O$_3$ & 0.62 & 0.69 &  &  &  \\ 
glycerol & 0.64 & 0.69 &  &  &  \\ \hline
Ca$_{0.4}$K$_{0.6}$(NO$_3$)$_{1.4}$ (CKN) & 0.54 & 0.70 &  &  &  \\ \hline
poly(propylene glycol) (PPG) & 0.46 & 0.77 &  &  &  \\ 
poly(ethylene terephthalate) (PET) & 0.54 & 0.76 &  &  & 
\end{tabular}
\caption{The scaling exponents $\Delta $ and $\protect\sigma $ for different
supercooled liquids. These exponents were obtained assuming that the MCT
predictions are valid.}
\label{tab:2}
\end{table}

\end{document}